\documentclass{aa}
\usepackage{graphicx}
\usepackage{txfonts}
\def\approxgt{\mathrel{\hbox{\rlap{\lower.55ex \hbox {$\sim$}}
        \kern-.3em \raise.4ex \hbox{$>$}}}}
\def\approxlt{\mathrel{\hbox{\rlap{\lower.55ex \hbox {$\sim$}}
        \kern-.3em \raise.4ex \hbox{$<$}}}}

\usepackage{natbib}
\def\chandra {\emph{Chandra }}
\def\chandran {\emph{Chandra}}
\def\xmmn {\emph{XMM-Newton }}

\def\xmms {\emph{XMM }}

\begin{document}
   \title{A cluster in a crowded environment: \xmmn and \chandra observations of  A3558}
   \titlerunning{ \xmmn and \chandra observations of A3558}
 \author{Mariachiara Rossetti
   \inst{1,2}
   \and
    Simona Ghizzardi
   \inst{2}
   \and
    Silvano Molendi
   \inst{2}
   \and
   Alexis Finoguenov
   \inst{3}
          }

\institute{
          Universit\`a degli Studi di Milano, Dip. di Fisica, via Celoria 16
         I-20133 Milano, Italy
         \and
         Istituto di Fisica Cosmica, CNR, via Bassini 15,
         I-20133 Milano, Italy
         \and
	 Max Planck Institut fur extraterrestrische Physik, D-85748 Garching, Germany
         }
\abstract{ Combining \xmmn and \chandra data, we have performed a detailed study of Abell 3558. Our analysis shows that its dynamical history  is more complicated than previously thought. We have found some traits typical of cool core clusters (surface brightness peaked at the center, peaked metal abundance profile) and others that are more common in merging clusters, like deviations from spherical symmetry in the thermodynamic quantities  of the ICM. This last result has been achieved with a new technique for deriving temperature maps from images.
We have also detected a cold front and, with the combined use of \xmmn and \chandran, we have characterized its properties, such as the speed and the metal abundance profile across the edge. 
This cold front is probably due to the sloshing of the core, induced by the perturbation of the gravitational potential associated with a past merger. The hydrodynamic processes related to this perturbation have presumably produced a tail of lower entropy, higher pressure and  metal rich ICM, which extends behind the cold front for $~ 500$ kpc.
The unique characteristics of A3558 are probably due to the very peculiar environment in which it is located: the core of the Shapley supercluster.

   }
\maketitle
\section{Introduction}
Superclusters are the most massive objects in the Universe and are
destined to collapse. By studying superclusters, we can observe the processes
related to individual components at an early stage of merging. These
are: mutual interaction between the components, accretion history of
the member clusters, interaction between the collapsed objects and the
intra supercluster medium. \\
The Shapley Supercluster \citep{shapley30} is one of the largest concentrations of mass in the local Universe \citep{bardelli94, zucca93, ettori97}.
The core of the concentration is dominated by the greatest cluster complex of the supercluster, which is composed of  A3562, SC 1329-313, SC 1327-312, A3558 and A3556 (from SE to W). Many multiwavelength observations have been carried out on the core of the Shapley Supercluster, from which the total gravitating mass is estimated to be $10^{15}-10^{16}\, h_{50}^{-1}\, M_{\odot}$ \citep{metcalfe94, ettori97}. The detected low surface brightness X-ray emission connects A3558 with the two groups and A3562 and probably also with A3556 \citep{kull99}. Also the galaxy isodensity contours extend over the whole complex \citep{bardelli98b}, indicating the presence of a structure on supercluster scales. \citet{akimoto03} have calculated the mean density of the core region of the Shapley supercluster to be 25 times greater than the critical density. This density is larger than the value needed to decouple from the cosmic expansion and contract, even if smaller than that needed to virialize. Therefore, they suggest that the whole A3558 complex is now in an early phase of contraction.  \\
The central galaxy cluster in the core region is A3558, which is also the brightest and most massive cluster in the complex. The ROSAT image of the core complex \citep{kull99} shows that the X-ray surface brightness of A3558 is elongated in the SE-NW direction, along which the other members and the galaxy distribution are also aligned \citep{bardelli94}. Therefore the infall of matter onto A3558  should be dominant along this direction.\\
The temperature distribution of the ICM has been investigated using ASCA data by \citet{mark97}, \citet{hanami99} and \citet{akimoto03}, showing deviations from spherical symmetry.\\
A radio analysis of the A3558 complex \citep{venturi00} underlines the absence of extended radio sources (halos and relics, which are observed in many merging clusters) in A3558. Also the radio-optical luminosity function for early-type galaxies in the complex is significantly lower than that derived for a sample of early-type galaxies in clusters.\\ 
In this paper we present results from \xmmn and \chandra observations of A3558. In Section 2 we present the analysis and  describe a new technique to produce temperature maps;  in section 3 we present the most interesting features of the cluster and in section 4 we summarize our results and discuss the dynamical state of the cluster.\\
At the nominal redshift of A3558 ($z=0.047$), 1 arcmin corresponds to 54 kpc ($H_0=70h_{70}\, \rm{km}\,\rm{s}^{-1}\,\rm{Mpc}^{-1}$). All quoted errors are at $1\sigma$, unless otherwise stated.

\section{Data analysis}
\subsection{XMM-Newton data reduction}
A3558 was observed by \xmmn during revolution 388 with the EPIC detectors in Full Frame Mode with the thick filter. 
We have obtained calibrated event files for the MOS1, MOS2 and PN cameras with SAS version 6.0. These files are then manually screened to remove remaining bright pixels or columns.
To reject  soft proton flares, we accumulate  light curves in the 10-12 keV energy band, where the emission is dominated by the particle-induced background. A visual inspection of the light curve reveals that the observation is not badly affected by soft protons: we remove only 0.5 ks, using the $\sigma$-clipping technique \citep{marty02}, leaving an exposure of 43 ks for the MOS cameras and 35 ks for the PN.\\For the study of galaxy clusters, especially in their outskirts, it is very important to subtract the instrumental background. The quiescent particle induced background can be properly subtracted  using a large collection of background data, such as long observations of blank sky fields. In this case, the choice was not easy because there are not many blank sky fields observations performed with the thick filter. We have selected two fields PHL 5200 \& PKS 0312-770 which have similar characteristics to the source (see Table \ref{obs_bkg}). We have applied to these  observations the same selection criteria applied to the source, we removed point sources, reprojected the files to the observation coordinates and merged the fields with the SAS task \emph{merge}, for a total exposure of 79 ks for the PN and 81 ks for the MOS detectors.\\
\begin{table}
\begin{minipage}[t]{\columnwidth}
\caption{Source and background observations.}
\label{obs_bkg}
\centering
\renewcommand{\footnoterule}{} 
\begin{tabular}{c c c}
\hline
\hline
Obj name &  $\rm{N_H}$ & SXR bkg \footnote{Average X-ray background  count rate in the 3/4 keV band, derived from the \emph{ROSAT} all-sky survey diffuse background maps.}  \\ 
 & ($10^{22}$cm$^{-2}$)&(cts s$^{-1}$arcmin$^{-2}$)\\
\hline
A3558 &  0.0389 & $(301 \pm 10)10^{-6}$\\
PHL 5200 & 0.0526 & $(111 \pm 8)10^{-6}$ \\
PKS 0312-770 & 0.0797 & $(122 \pm 6)10^{-6}$\\
\hline
\end{tabular}
\end{minipage}
\end{table}
The intensity of the quiescent particles background may differ during the observations of the source and of the blank fields. We have compared the count rates at energies greater than 10
keV, where the emission is dominated by the background,  in the source and blank sky fields: the level of count rates during background observations is about 10\% higher than that of the source observation. 
To correct for this effect we renormalized the background by multiplying it by a factor of 0.87 before performing background subtraction.

\subsection{\chandra data reduction}
\chandra observed A3558 in 2001 with the Advanced CCD Imaging Spectrometer (ACIS). We have created standard level two event files, screened for background flares using version 2.4 of the ACIS calibration products available with the version 3.02 of the CIAO package. 
The useful time for the analysis after the screening is 13.9 ks. \\
We have performed background subtraction using the ``blank sky'' datasets enclosed in the CIAO Calibration Database. We have processed the background file with the same procedure of the source file and we have reprojected it to the Abell 3558 observation sky coordinates.

\subsection{Maps from images}
Using data from X-ray imaging spectrometers, such as EPIC and ACIS, we can derive a 2-dimensional temperature map of the cluster. 
In order to exploit the spatial resolution of these instruments, we want to obtain at the same time as much ``spectral'' and ``spatial'' information as possible.
To do this, we have derived temperature maps from X-ray images using two different and complementary techniques:  the wavelets + hardness ratio analysis \citep{briel04} and the broad band fitting with adaptive binning, that we will describe in section \ref{technique}. Combining the temperature map and the surface brightness image, we also derived pseudo-pressure and pseudo-entropy maps. 
Besides giving informations themselves, the maps can be used to select interesting regions for a proper spectral analysis.

\subsubsection{Maps from adaptive binning + broad band fitting}
\label{technique}
   \begin{figure*}
    \vspace{-2.5 cm} 
   \hspace {-2.5 cm}
   \includegraphics[angle=90, width=22 cm]{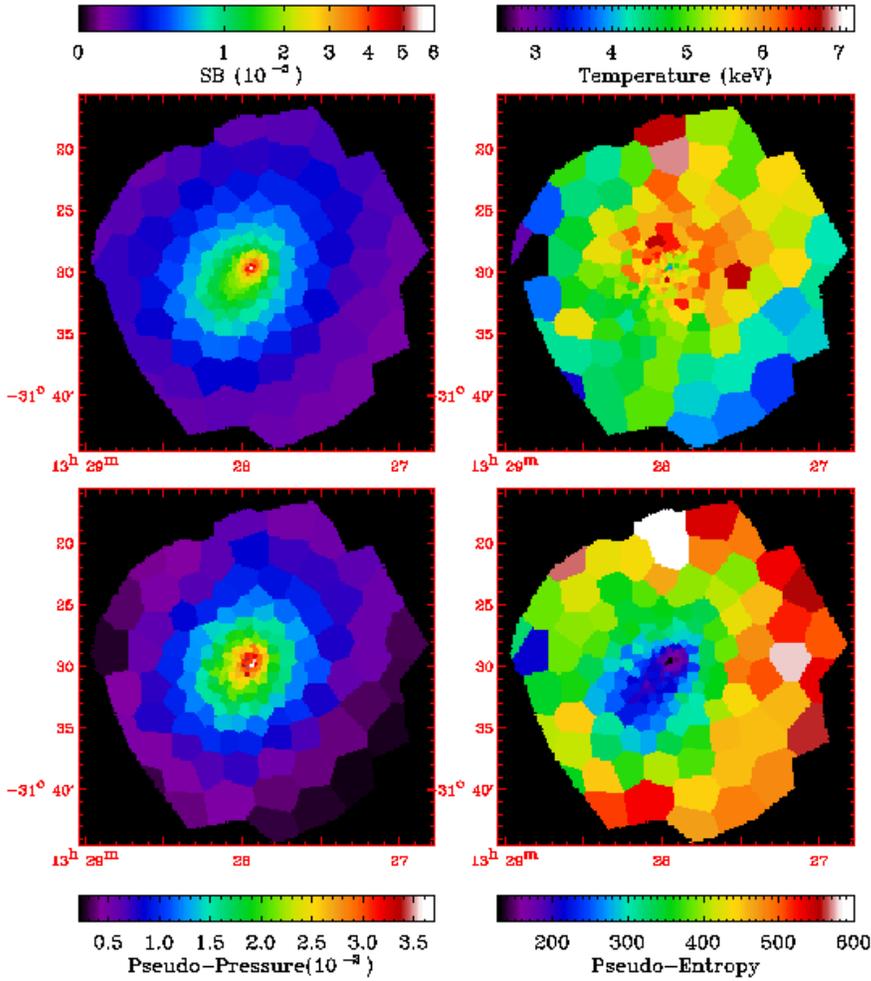}
   \caption{Maps of the main thermodynamic quantities of the cluster, obtained with the adaptive binning and broad band fitting technique. \emph{Upper left:} ``projected emission measure'' (XSPEC Normalization per pixel); \emph{Upper right:} temperature (keV), \emph{Lower left:} pseudo-pressure (arbitrary units); \emph{Lower right:} pseudo entropy (arbitrary units). Coordinates on the maps are right ascension and declination. }
              \label{map_ad}%
    \end{figure*}

Temperature maps of galaxy clusters are often obtained applying adaptive smoothing to X-ray images \citep{mark00}, in order to emphasize spatial details. The smoothing technique increases the local signal to noise ratio by correlating neighboring pixels through a convolution process. One of the problems of this technique is that it is often difficult to handle the statistics of the  correlations introduced by the smoothing. The adaptive binning technique  locally groups and averages together data, allowing us to have a firmer  control on the statistics:  this is why it should be preferred when a precise error estimation is needed. \\ 
Rather than using the hardness ratio technique, we have performed a ``broad band fitting'' starting from images, as in \citet{mark00}. The combined use of these techniques allows us to have a good control on the statistics and to produce quantitative thermodynamic maps (i.e. maps with their associated errors). In the following, we will describe the application of this technique to EPIC images, but it can also be easily applied  to ACIS data (see \citealt{ghizza05} for an application to \chandra images).\\
We extract images in five energy bands (0.4-0.8 keV, 0.8-1.2 keV, 1.2-2 keV, 2-4 keV, 4-10 keV) from the event files of the observation and of the background and we generate exposure maps in the same energy bands with the SAS task \emph{eexpmap}. Then we select the images with lowest signal (i.e. one of the MOS images in the hardest energy band) and we perform adaptive binning on this image, in order to group pixels in bins with the same signal to noise ratio. To do this we have used the 2-dimensional adaptive binning technique by \citet{cappellari}, which is based on the Voronoi tessellation.
This procedure has been applied to the MOS1 image of A3558 in the 4-10 keV  energy band with a target signal to noise ratio of 10; the same binning is applied to all images, assuring a minimum S/N=10 in all bins in all images, since the statistics in the PN and in the other bands is better than in the image we have selected.
For each bin, we sum the count rates of MOS1, MOS2 and PN, obtaining the EPIC count rate and its error in each band and we compare them with a grid of predicted count rates, generated with XSPEC using absorbed \emph{mekal} models with different temperatures and normalizations, while the absorption column and redshift are fixed to the known values for this cluster and the iron abundance is fixed at  $0.3$ solar.
We select our ``best fit'' values for temperature and normalization as those that minimize $\chi ^2$. Using the $\chi ^2$  matrix, we can also find values  within a fixed $\Delta\, \chi ^2$ and we can associate errors with  best fit parameters.\\
The normalization values are in XSPEC units:
\begin{equation}
K= \frac{10^{-14}}{4 \pi D_A(1+z)^2} \int n_e n_H dV,
\end{equation}
where $D_A$ is the angular size distance to the source (cm), $n_e$ is the electron density (cm$^{-3}$), and $n_H$ is the hydrogen density (cm$^{-3}$). If we divide the normalization by the number of pixels in each bin ($K/N_{p}$), we obtain the ``projected emission measure'' that is proportional to the square of the electron density $n_e^2$, integrated along the line of sight.   \\ 
Using this ``projected emission measure'' we can obtain for each bin the ``pseudo-entropy'' $s$ and ``pseudo-pressure'' $P$, defined as:
\begin{equation}
s \equiv T\left( \frac{K}{N_p} \right)^{-1/3}\\
P \equiv T\left( \frac{K}{N_p} \right)^{1/2},
\end{equation}
and we can produce maps of these important thermodynamic quantities. \\  
The maps obtained with this technique with EPIC are shown in Fig. \ref{map_ad}. Our technique can also produce for each of these maps, two maps corresponding to the upper and lower limit at a given confidence level. These maps are extremely useful to determine whether the gradients that we observe in Fig. \ref{map_ad} are significant or not.
\begin{figure}
  \vspace{-2.5 cm}
     \hspace{-1.2 cm}
 \includegraphics[angle=90]{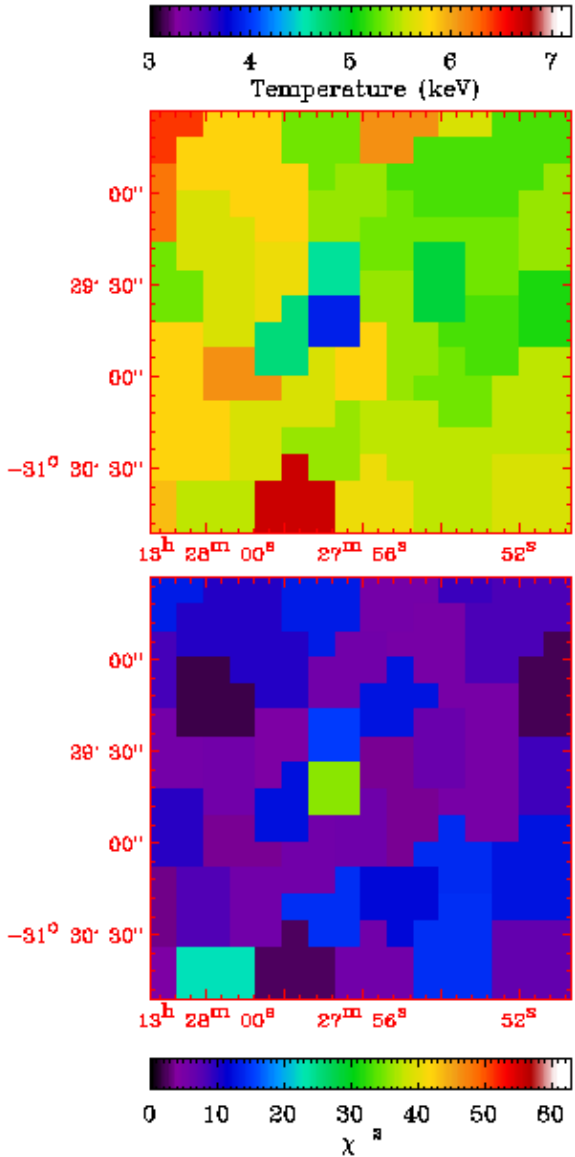}
     \caption{Center of A3558 (inner $~160$ arcsec). \emph{Upper panel}: Temperature map, \emph{lower panel}: $\chi^2$ map (3 d.o.f.). Coordinates on the maps are right ascension and declination.}
              \label{center}%
    \end{figure}
We have applied this analysis to the ``blue spot'' in the temperature map at the center of the cluster (Fig. \ref{center}a), corresponding to the brightest bin. Its temperature, kT$=(3.7^{+0.22}_{-0.18})$ keV, is significantly lower than the mean value of  the adjacent bins, kT=$(4.89 \pm 0.11)$ keV.
Moreover the $\chi ^2$ of the ``fit'' in this bin is higher than in the surrounding bins (Fig. \ref{center}b), suggesting that the fit in the blue spot is worse than in the other regions. Investigation of the \chandra image (Fig.\ref{cha_im}) shows that the blue spot coincides with the AGN located in the BCG. This shows that our technique is able to recognize the spectral shape, even if we use large energy bands.  \\
The typical relative errors on the temperature  are of the order of $3-5 \%$ in a circle of radius 7 arcminutes from the SB peak, while at outer radii, errors grow to about $5-10 \%$. The larger errors ($\sim 10 \%$) are found in the southern and western regions, where the surface brightness decreases more rapidly: in these regions the dominant factor of indetermination in temperature is the background subtraction. \\
We do not show here the maps obtained from ACIS images: because of the short exposure of this observation we cannot produce  maps with bins smaller than in EPIC maps. 
A deeper \chandra observation could allow us to exploit the full potential and the spatial resolution of the ACIS instrument.

\subsubsection{Maps from wavelet + hardness ratio} 

   \begin{figure*}
     \vspace{-2.5 cm}
     \hspace{-2.5 cm}
    \includegraphics[angle=90, width=22 cm]{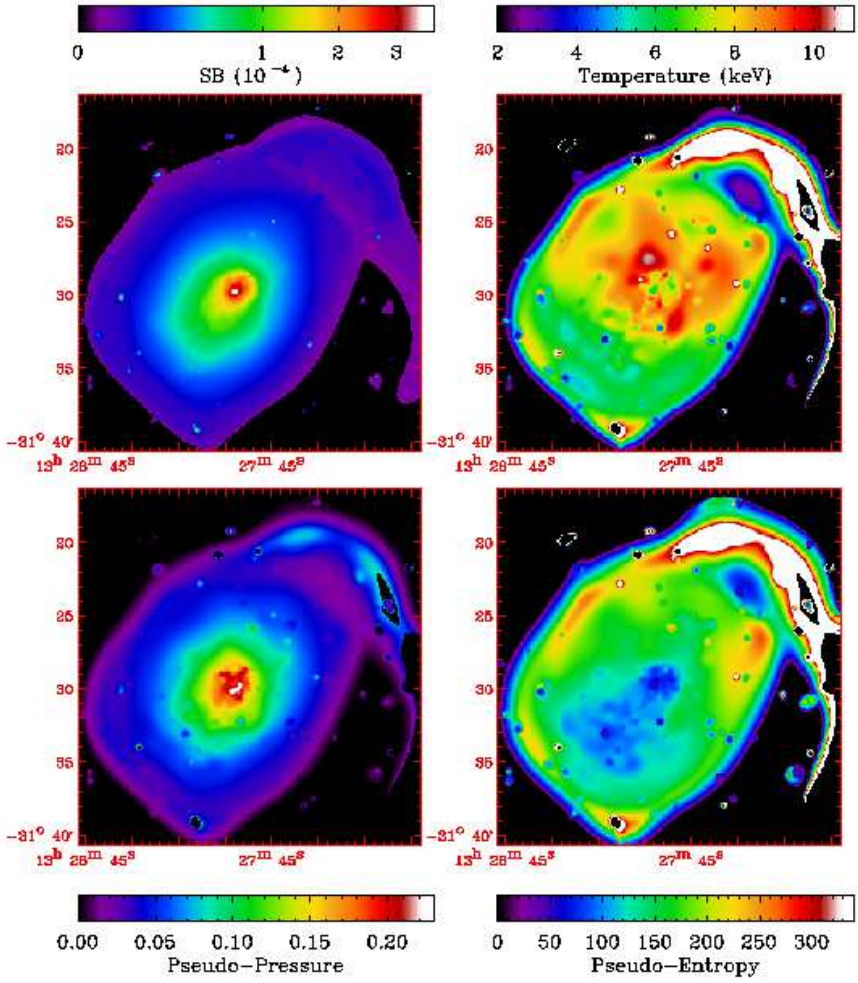}
     \caption{Maps of the main thermodynamic quantities of the cluster, obtained with wavelet decomposition and hardness ratio.\emph{Upper left:} wavelet decomposed image of the cluster; \emph{Upper right:} temperature (keV), \emph{Lower left:} pseudo-pressure (arbitrary units); \emph{Lower right:} pseudo-entropy (arbitrary units). Coordinates on the maps are right ascension and declination. The hot zone in the NW is an artefact due to both high background in the hard band and to its position near  the edge of the FoV.}
              \label{map_al}%
    \end{figure*}
In Fig. \ref{map_al} we show the maps of the main thermodynamic quantities obtained with hardness ratio (HR) and wavelet decomposition, with the technique described in Briel et al. (2003). The interest in this technique is more focused on detecting the regions where there are temperature gradients than in the absolute values, since they are used to select  regions for a proper spectral analysis. The main advantage of the wavelet decomposition is to reveal the small-scale features, which may not be resolved by the binning method.\\
A qualitative comparison with the maps in Fig. \ref{map_ad} shows that they are consistent: the features detected with the new technique are seen also in Fig. \ref{map_al}.\\
Only the temperature maps can be compared quantitatively (Fig.~ \ref{map_ad}b and Fig.~ \ref{map_al}b), since the other maps have different units. 
 The fact that temperatures are always higher in the HR map is not worrying because the purpose of this technique is not to produce quantitative temperature maps.
Moreover, the simplified HR technique that we have used here assumes that the observation is performed with a medium filter and therefore can overestimate the temperature.

\subsection{Maps from spectral analysis}

\begin{figure}
   \centering
   \includegraphics[angle=-90,width=8cm]{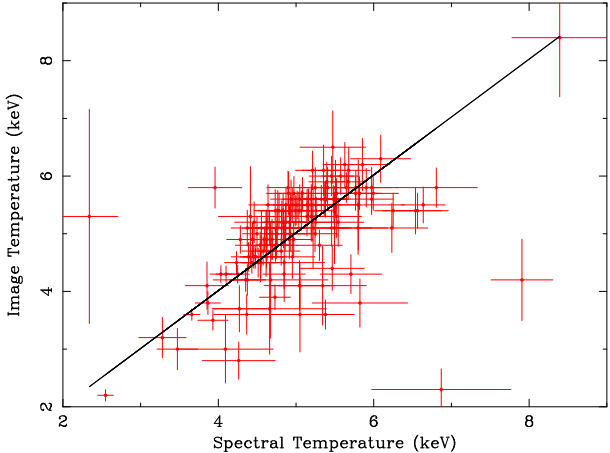}
   \caption{Temperature derived in each polygon with our technique versus the weighted mean of the best fit temperature in each detector obtained from spectral analysis. The continuous line represents the ideal expected relation $T_{image}=T_{spectral}$.}
   \label{confrontot}%
\end{figure}

In order to test our technique we have compared the temperature and normalization maps that we derive from images with the same maps obtained from a proper spectral analysis. 
With the results of the spectral analysis we have produced temperature and ``projected emission measure'' maps that can be compared to the maps obtained from images (Fig.\ref{map_ad}). \\
A plot of the temperature obtained with our technique versus the temperature obtained with spectral analysis   is shown in Fig. \ref{confrontot}.\\
The mean absolute deviation ($|T_{image}/T_{spectral}-1|$)  is $~8.8\%$, with a $1.7\%$ dispersion around the mean. The largest deviation between the temperature estimates is $(T_{im}-T_{sp})=4.5(\sigma_{im}^2+\sigma_{spe}^2)^{0.5}$).  A similar analysis of the difference between the temperature obtained using MOS1 data and that obtained from the PN shows comparable variations (mean deviation $17.3\%$ and scatter $ 5.3\%$).\\
Therefore the  error associated with our technique is somewhat smaller than the systematic error associated with cross-calibration uncertainties between MOS and PN.\\
The good agreement between the results of our technique and those of the spectral analysis confirms that the maps obtained from images can substitute for those obtained from the spectra in all the cases where a proper spectral analysis would require too much time, as for instance in long observations of bright clusters or in large samples.  

\subsection{Ratio maps}
\begin{figure*}
   \centering
   \includegraphics[width=11 cm, angle=90]{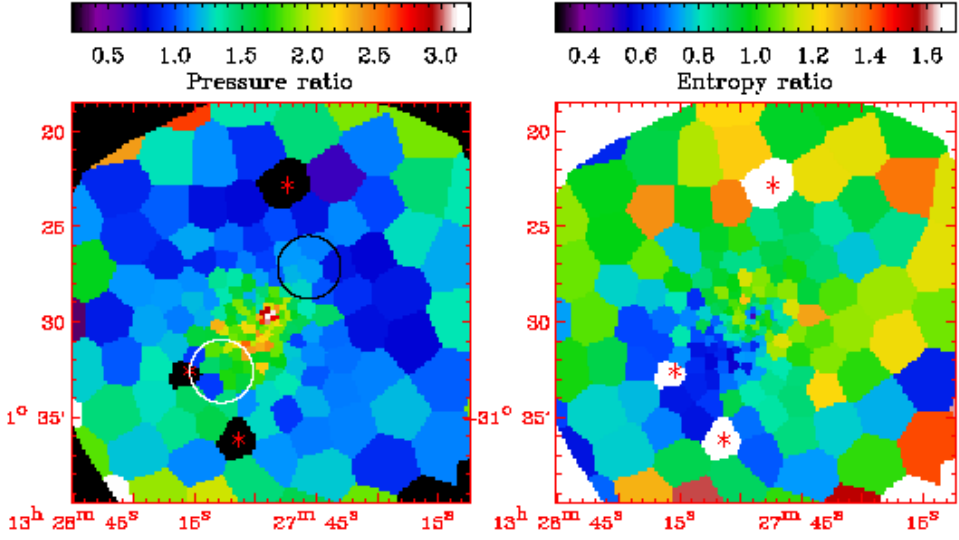}
   \caption{Ratio maps of pressure (left) and entropy (right). The regions labeled ``*'' have no data because of the subtraction of point sources. The white circle is the region considered for the analysis of the pressure excess, while the black circle is the reference region.Coordinates on the maps are right ascension and declination. }
              \label{ratios}%
    \end{figure*}

The most interesting features in a non-relaxed galaxy cluster are the deviations from the spherical symmetry, which would be expected in an ideal relaxed cluster. 
Such deviations can be seen in any  ICM observable (see for instance the ``residual'' image of the Coma Cluster, by \citealt{neumann03}). We have decided to look for deviations in pressure (since it is strongly related to the Dark Matter distribution) and in entropy. 
First we have derived pressure and entropy in all the polygons produced by the Voronoi tessellation, using the results of spectral analysis and making the volume estimates as described in \citet{henry04} and \citet{mahd05}. Then we calculate the predicted ``relaxed'' values, using the  best fit models  to the pressure and entropy profiles in  a sample of clusters \citep{fino05} and centering them on the surface brightness peak (the impact of the choice of the center of symmetry will be discussed later). Finally we divide the observed values by the predicted one in each polygon,  obtaining the ratio maps shown in  Fig. \ref{ratios}.
The area-weighted fractional scatter in the entropy and
pressure is $0.23\pm0.01$ and $0.27\pm0.01$, respectively, inside a circular region of $r=0.5r_{500}$ ($r \simeq 9.2$ arcmin from the cluster center).\\
Since this technique is strongly based on geometrical and symmetry assumptions, we have recalculated the maps using another point as the center of the cluster. This ``new center'' is the center of the surface brightness isophotes (we will discuss in Sec. \ref{results} the choice of this point) and is located $\sim 70$ arcsec SE of the surface brightness peak. In Table \ref{tab2}, we compare the fractional scatter obtained using the different centers.\\
The choice of the center does not have a strong impact either on the values of the pressure and entropy scatter, especially at large scales,  or on the ratio maps. The structures that appear in Fig. \ref{ratios} are seen also in the maps with the other center (that we do not show here). Some obvious changes can be found only in the inner regions: if we use the ``SB center'' the main surface brightness peak appears as a substructure.
 
\begin{table}
\begin{minipage}[t]{\columnwidth}
\caption{Fractional entropy and pressure scatter obtained using the surface brightness peak (SB peak) and the center of large scale X-ray isophotes (SB center) as center of symmetry.}
\label{tab2}
\centering
\begin{tabular}{c c c c}
\hline
\hline
Region \footnote{Circular region inside which the scatter is calculated. The radius is expressed in fraction of $r_{500}$: radius inside which the  cluster mass density is 500 times the critical density of the universe.} & center & Entropy & Pressure  \\
\hline
$r<0.1 r_{500}$ & SB peak & $(0.179 \pm 0.011)$ & $(0.629 \pm 0.011)$ \\
                & SB center & $(0.193 \pm 0.008)$ & $(0.667 \pm 0.012)$ \\
\hline
$r<0.2 r_{500}$ & SB peak & $(0.180 \pm 0.006)$ & $(0.413 \pm 0.006)$ \\
                & SB center & $(0.161 \pm 0.005)$ & $(0.448 \pm 0.007)$ \\
\hline
$r<0.5 r_{500}$ & SB peak & $(0.227 \pm 0.010)$ & $(0.276 \pm 0.009)$ \\
                & SB center & $(0.233 \pm 0.012)$ & $(0.272 \pm 0.007)$ \\
\hline
\end{tabular}
\end{minipage}
\end{table}

\section{Results}
\label{results}
\begin{figure}
  \vspace{-2.5 cm}
  \hspace{-1. cm}
   \includegraphics[width=12 cm, angle= 90]{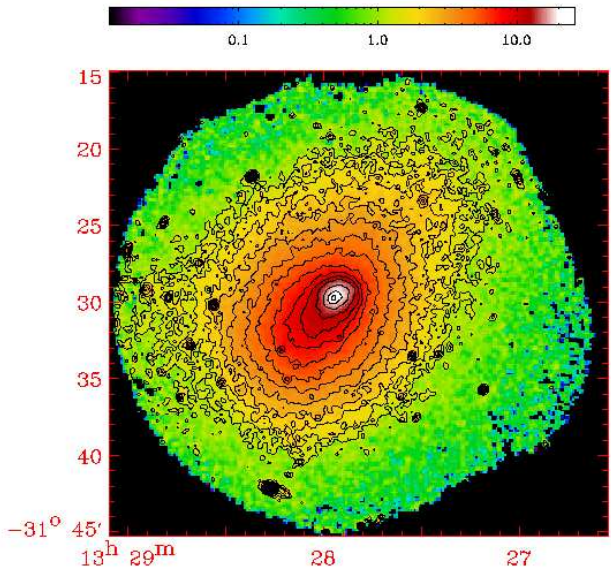}
   \caption{This EPIC flux image in the 0.4-2 keV  band is computed following a procedure similar to the one described in \citet{baldi02}. 
   The EPIC source counts image, $S_{EPIC}$, is computed by summing the MOS1,
MOS2 and PN source images, while  
   an EPIC source exposure map, $t_{EPIC}$, is computed by summing the  source
 exposure maps of each detector.  
  The EPIC count rate image is defined as: $ cr_{EPIC} \equiv S_{EPIC} / t_{EPIC}$.
   The total conversion factor (count rate to flux) $cf_{EPIC}$ has been calculated using the
exposure times for MOS1, 
  MOS2 and PN, the conversion factors for the three instruments,
$cf_{MOS1}$, $cf_{MOS2}$, $cf_{PN}$,    and following the formula:
   $ {t_{EPIC} \over cf_{EPIC}} = {t_{MOS1} \over cf_{MOS1}} + {t_{MOS2}
\over cf_{MOS2}} + {t_{PN} \over cf_{PN}} $.
 It is easy to show that the EPIC source flux image is then : $ F_{EPIC} =
cf_{EPIC} \cdot cr_{EPIC} $.
  Similarly we compute the EPIC background flux image from the MOS1, MOS2
and PN background images and exposure maps. 
  Finally by subtracting the EPIC background flux image from the source
flux image we derive an EPIC net flux image in units of $10^{-15}\rm{ergs}\ \rm{cm^{-2}}\ \rm{s^{-1}} \rm{pixel^{-1}}$ (one pixel is $8.5*8.5 \rm{arcsec}^2$) .
Coordinates on the image are right ascension and declination.}
              \label{map_epic}%
    \end{figure}

The morphology of the cluster in  X-rays  can provide important information on its dynamical state.
In Fig. \ref{map_epic} and \ref{cha_im} we show the images from EPIC and ACIS-S3 respectively.
The peak of the surface brightness coincides with the BCG, which hosts an AGN.  Comparing the image of this cluster with images from a sample of nearby clusters \citep{ghizza05} we have observed that the X-ray emission is not as peaked as in cool core clusters, but the peak is more apparent than in the clusters that are currently undergoing a major merger. \\
The surface brightness contours in Fig. \ref{map_epic} show that the distribution of the ICM is elliptical, elongated in the SE-NW direction (inclination$=50^{\circ}$).
Excluding the inner regions ($r < 120$ arcsec from the SB peak), the cluster appears remarkably symmetric large scales. Approximating the contours in Fig.\ref{map_epic} with ellipses, we found that their center does not change significantly if we exclude the inner contours. This center can be considered the center of symmetry of the cluster: it is located $\simeq 70$ arcsec SE of the surface brightness peak. Such offsets are not uncommon in clusters. \\
In the \chandra image (Fig. \ref{cha_im}) there is a clear brightness discontinuity (visible even if at a lower resolution in Fig.\ref{map_epic})  in the NW sector. This edge is similar to those observed with \chandra in many clusters \citep{mark00, vikh01b, dupkewhite}: the analysis of the surface brightness profile and of the temperature discontinuity (Section \ref{coldfront}) confirms that it is a cold front. \\
The temperature maps (Fig. \ref{map_ad}b and Fig. \ref{map_al}b) confirm that the cluster is not spherically symmetric: we see a ``hot bridge'' ($kT = (6.2 \pm 0.5)$ keV) which surrounds the cluster center on three sides, while the temperature in the SE sector at the same distance from the center is  significantly lower ($kT = (5.1 \pm 0.4)$ keV, $\sim 20\%$), similar to the temperature of the cluster core. The entropy maps (Fig. \ref{map_ad}d and Fig. \ref{map_al}d) do not show the spherical symmetry typical of relaxed clusters: the low entropy gas extends from the core towards the SE for about 500 kpc. 
The pressure maps (Fig. \ref{map_ad}c and Fig. \ref{map_al}c) seem to suggest that this quantity is more spherically symmetric than the others but a detailed analysis shows that there is an elongation of contours in the direction of the low-entropy tail. This is also confirmed by the ratio pressure map (Fig. \ref{ratios}) which shows a clear deviation from radial symmetry in the SE sector and the presence of a secondary pressure peak. The entropy ratio map (Fig. \ref{ratios})  confirms the presence of low-entropy gas in the SE sector. These features in the ratio maps are still present even if we select the large scale center of symmetry as our center.\\
It is apparent in our analysis that the regions where the ICM shows the most deviant features from the rest of the cluster are located in the SE sector.\\
In comparison to the typical
scatter revealed in a representative sample (Finoguenov et al. 2005),
A3558 does not exhibit an anomalous amount of entropy or pressure substructures at large scales ($r<0.5r_{500}$).

\subsection{The cold front}
\label{coldfront}
\begin{figure}
  \vspace{-2.5 cm}
  \hspace{-0.5 cm}
   \includegraphics[width=12 cm, angle=90]{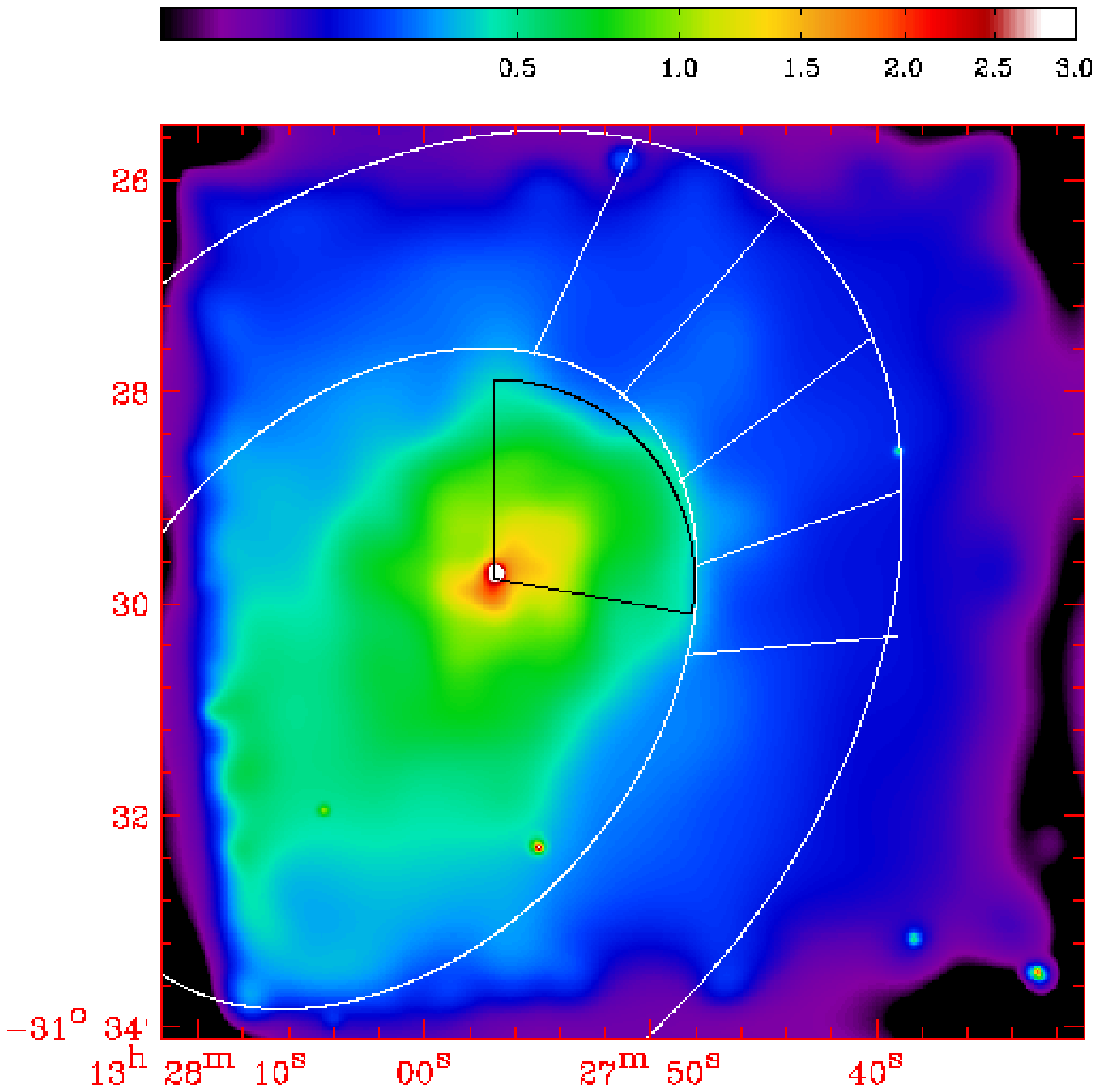}
\caption{\chandra image of the cluster in the band $0.8-3$ keV, adaptively smoothed using the CIAO task \emph{csmooth}. The black curved line shows the position of the cold front, and the white elliptical sectors are used for the surface brightness profiles in Appendix A. Coordinates on the image are right ascension and declination.}
          \label{cha_im}%
    \end{figure}

\begin{figure*}
 \centering
   \includegraphics[ width=16 cm ]{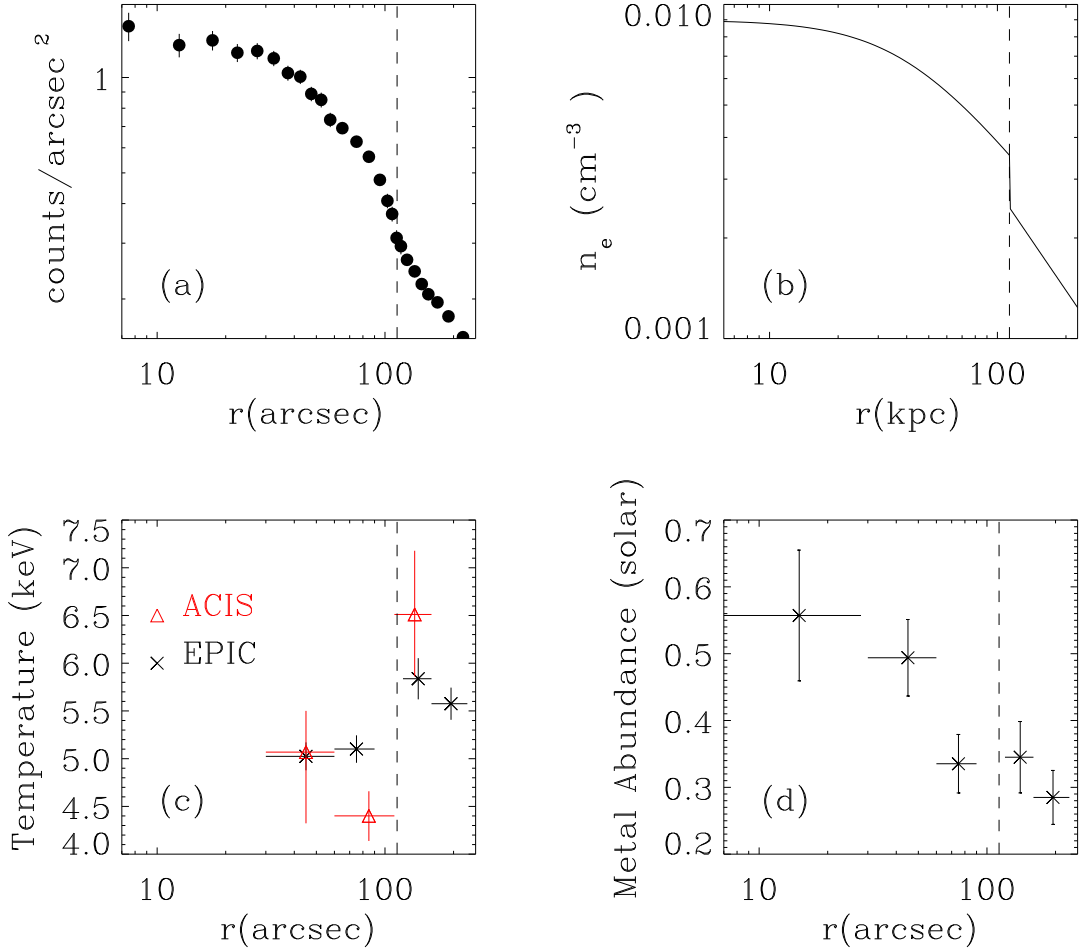}
     \caption{\emph{Upper left}: Surface brightness profile obtained with \chandra across the discontinuity. \emph{Upper right}: Electron density profile with the model described in Appendix A.
\emph{Lower left}: Projected temperature profiles across the cold front, obtained from ACIS spectra (red triangles) and EPIC spectra (black crosses). In order to reduce the effect of the EPIC PSF we did not  extract the spectra in a slice of width 30 arcsec around the cold front. \emph{Lower right}:  Metal abundance profile across the cold front, using \xmmn spectra.}
      \label{cf_fig}
  \end{figure*}

In order to characterize the edge seen in \chandra images, we have extracted the surface brightness profile across the edge  (Fig. \ref{cf_fig}a): the discontinuity is apparent from the change of the slope at  a projected distance $b \simeq 100$ arcsec from the cluster center.\\
We have extracted spectra from \chandra and \xmms data in circular annuli inside and outside the discontinuity. Fitting them with mekal models, we derive the  temperature profile across the edge (Fig. \ref{cf_fig}c): the gas is colder where the surface brightness is higher and therefore it is a cold front. Both instruments show the temperature jump across the discontinuity, but  with the better statistics of the \xmmn observation, the temperature values measured by EPIC are more accurate.\\
Measuring the pressure jump across the cold front allows us to calculate the speed of the moving cloud \citep{landau59, vikh01b}:
\begin{displaymath}
\frac{P_{st}}{P_{out}}=\left\{ \begin{array}{ll}
\left(1+\frac{\gamma-1}{2}M^2\right)^{\gamma/(\gamma-1)} & M<1 \\
M^2\left(\frac{\gamma+1}{2}\right)^{\frac{\gamma+1}{\gamma-1}}\left(\gamma -\frac{\gamma-1}{2M^2}\right)^{-1/\left(\gamma-1\right)} & M>1 \end{array}\right.
\end{displaymath}
where $P_{st}$ is the pressure at the stagnation point, that can be approximated by  the pressure just inside the cold front, $P_{out}$ is the pressure of the external, undisturbed gas, $\gamma$ is the adiabatic index and $M$ is the Mach number of the motion of the cold cloud into the external gas. 
To calculate the pressure ratio, we derive the density jump from the fit of the surface brightness profile with the model described in Appendix A: we obtain $n_{in}/n_{out}=1.88 \pm 0.13$. We calculate the temperature jump with \xmmn since the statistical errors on the temperatures are smaller. After deprojection \citep{ettori02}, we obtain $T_{in}/T_{out}=0.85 \pm 0.11$. This leads to a pressure ratio $P_{st}/P_{out}=1.60 \pm  0.19$, that corresponds to a subsonic motion, with a Mach number $M=0.78_{-0.15}^{+0.12}$. 
Therefore, the cloud is moving with a speed $v=Mv_s=936_{-180}^{+144}\, \rm{km}\,\rm{s}^{-1}$, relative to the gas outside the cold front.\\
In the expression taken from \citet{landau59}, $P_{out}$ should be the pressure far upstream from the cold front, but unfortunately this does not apply easily to clusters since the pressure exhibits a strong gradient. For this reason, we have used for $P_{out}$ the value in a region immediately outside the edge, which could be compressed and heated by the motion of the cold cloud. Therefore, we must consider the Mach number $M=0.8$ a lower limit for the speed of the cloud.\\
If the motion of the cold front were supersonic, we would expect the presence of a shock in front of the moving cloud.
The shock would compress and heat the ICM leading to a second discontinuity in surface brightness and temperature. 
We have searched for the possible bow shock using the \xmms observation, analyzing surface brightness and temperature profile in elliptical annuli in the sector shown in Fig.\ref{cha_im}.
Up to 200 arcsec from the edge of the cold front, there are no significant indications of a change in the slope in the surface brightness profile or of a temperature discontinuity. We can use Moeckel's method \citep{vikh01b} to determine an upper limit to the Mach number of the moving cloud:  a stand-off distance between the cloud and the shock of $\simeq 200$ arcsec corresponds to a Mach number $M\simeq 1.09$. Therefore, we can conclude that the motion of the cloud is probably subsonic, with a Mach number in the range $0.78-1.09$. The large indetermination on this result reflects the importance of the unavoidable assumptions on which it is based. Also the geometry and the symmetry of the system are very important and to prove this, we have used a simpler and less accurate model than that in Appendix A to fit the surface brightness profile across the edge, assuming spherical symmetry centered on the BCG. With this model we have found a lower density jump ($n_{in}/n_{out}=1.43 \pm 0.04$) and therefore a lower but consistent Mach number  $M=0.53 \pm 0.07$.  \\
The metalicity profile across the cold front obtained with \xmmn spectra (Fig. \ref{cf_fig}d) could give us potentially important information on the origin of cold fronts. Large metalicity discontinuities are expected in some cold fronts due to merging where the core of an interacting subcluster is traveling through the ICM of a cluster poor in metals or through the outskirts of a cluster. The absence of a strong discontinuity in the metal abundance is consistent with several possible scenarios: a ``merging'' cold front which travels through an ICM that happens to have a similar metalicity and also a ``sloshing'' core \citep{mark01a}. \\
The mean metalicity inside the cold front, $Z_{in}=(0.41 \pm 0.03)\, Z_{\odot}$, is higher than the abundance in the gas outside, $Z_{out}=(0.31 \pm0.03)\, Z_{\odot}$, at a 3 $\sigma$ level, but the profile is consistent with a smooth decreasing profile, like those observed in   cool core clusters \citep{grandi01}, or with a metalicity jump less than 0.1 (30-40 \%). We conclude that the present data do not allow us to rule out any possible interpretation.\\ 
 \begin{figure}
  \vspace{-2.0 cm}
  \hspace{-1.0 cm}
   \includegraphics[width=12 cm, angle=90]{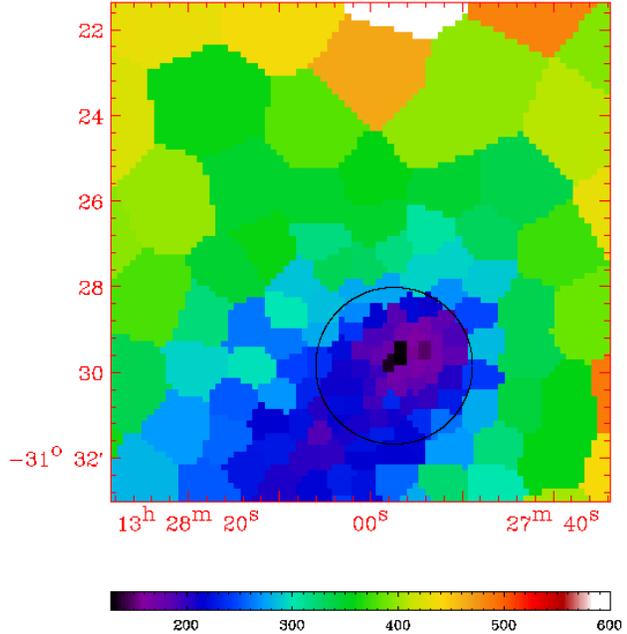}
\caption{Entropy map in the inner regions of the cluster. The circle radius is 113 arcsec, which corresponds to the distance between the center and the cold front. The gas with lowest entropy is located west from the center, in the direction of the surface brightness discontinuity. Coordinates on the image are right ascension and declination.}
\label{zoom_s}
\end{figure}
The distribution of entropy inside the core of the cluster (inner 120 arcsec) is intriguing: the gas with lowest entropy is located NW of the central AGN, near the discontinuity of the cold front (Fig. \ref{zoom_s}). In some numerical simulations \citep{heinz03, ascas06}, the flow of gas past the moving cloud induces slow motions inside the cloud, which should transport gas from the center of the subcluster  towards the interface. Our entropy distribution suggests that the low entropy gas located at the center of the cluster is moving towards the surface brightness discontinuity.  \\

\subsection{The anomalous SE sector}
\begin{figure}
\label{depro}
 \centering
\vbox{
   \includegraphics[width=6.5 cm, angle=90]{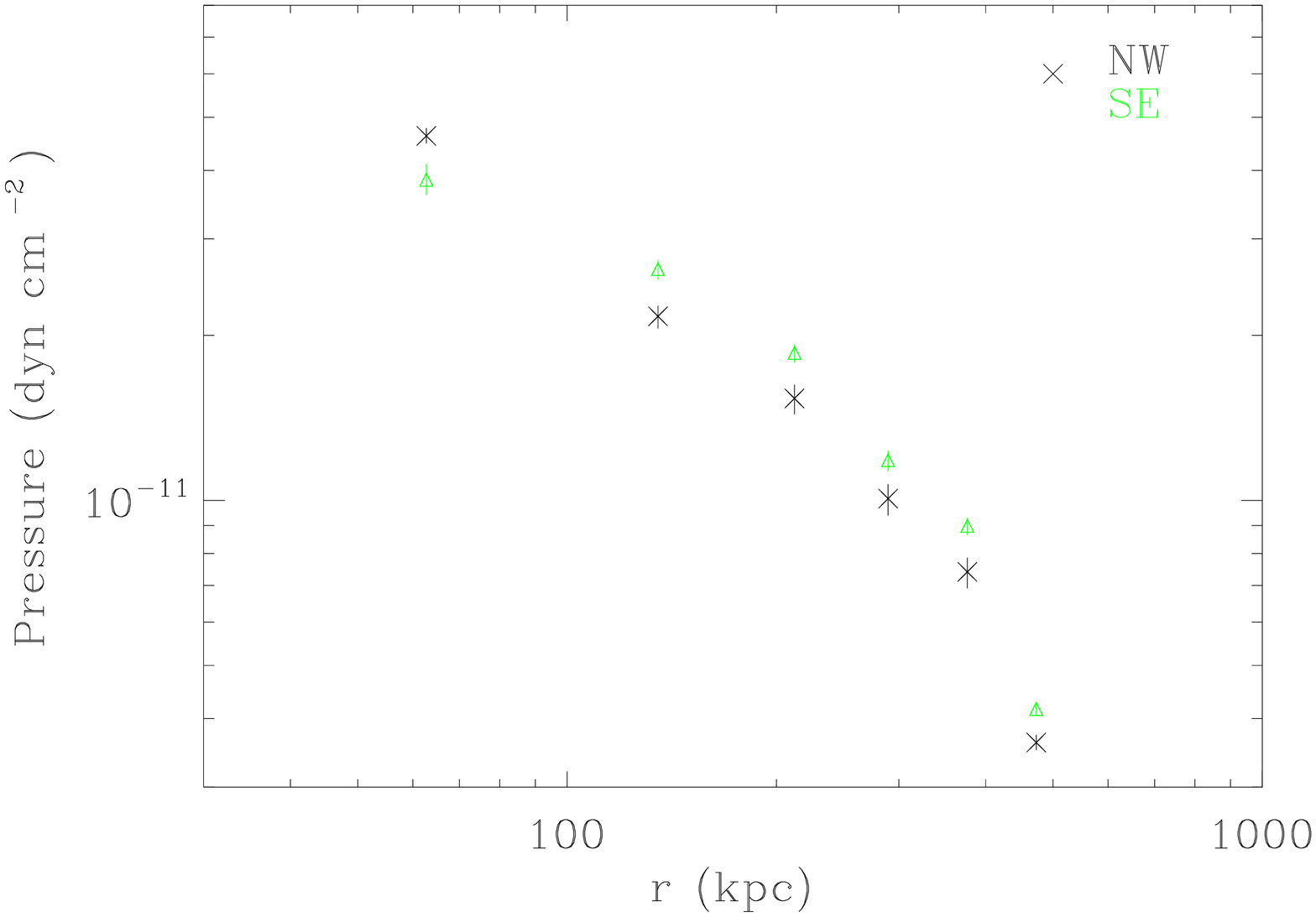}
    \includegraphics[width=6.5 cm, angle=90]{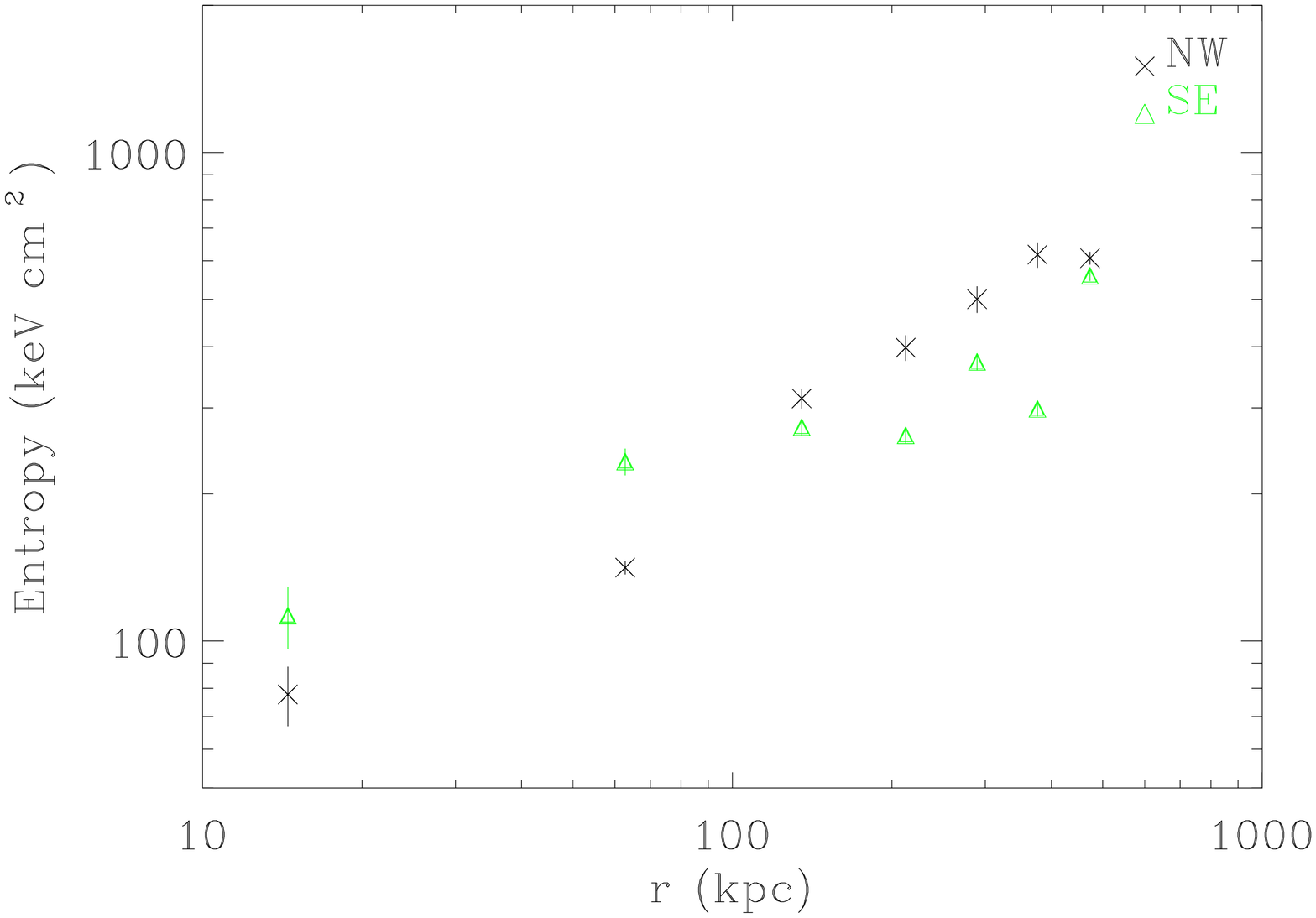}
}
\caption{Deprojected pressure (upper panel) and entropy (lower panel) profiles in the SE (green triangles) and NW (black crosses) sector, using \xmmn spectra.}
\end{figure}

As we have already outlined, there is a direction in A3558 in which the ICM seems to have thermodynamic properties different from what we found in other sectors of the cluster: the SE regions.
A low-entropy tail is observed in  Fig. \ref{map_ad}d and Fig. \ref{map_al}d, extending from the core toward the SE for $\sim 500$ kpc.
Also the pressure is different in this sector, as seen in the ratio map (Fig. \ref{ratios}): it is about a factor of two higher than predicted.\\
The presence of unresolved clumps of denser matter could bias our estimate of pressure, since the density fluctuations enhance the cluster luminosity over what would be expected for a smooth, single-phase ICM \citep{mathiesen99}.  From the test described in Appendix B, we conclude that the pressure excess is not an artifact due to unresolved clumps and that it must have a physical origin. \\ 
Projected entropy and pressure map are perhaps not completely reliable. To test the robustness of the detection of the anomalous thermodynamic properties of the SE sector we have compared these results to those obtained from an independent technique.
In Fig.\, \ref{depro} we show the deprojected pressure and entropy profile in two opposite sectors of the cluster, obtained with spectral analysis and the deprojection technique described in \citet{ettori02}. The entropy in the SE sector between 100 and 500 kpc is significantly lower than in the other direction, as predicted by the projected maps and the deprojected data are consistent with a pressure excess in these regions, even if at lower statistical significance. \\
\begin{figure}
 \centering
   \includegraphics[width=9 cm]{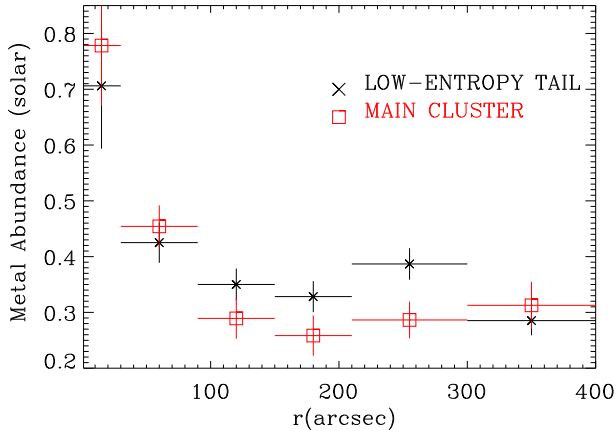}
\caption{Metal abundance profile in the low entropy tail (black diagonal crosses) and in the main cluster (red boxes) , using \xmmn spectra.}
\label{met_tail}
\end{figure} 
In order to study the metal distribution, we have extracted spectra from radial annuli in two sectors: one corresponding to the low entropy tail and the other to the rest of the cluster. These profiles (Fig. \ref{met_tail}) show that the abundance is  higher in the regions of the low entropy gas than in the other regions. This suggests that the chemical history of the gas is different in the SE region with respect to other regions of the cluster.\\
The anti-correlation between entropy and metal abundance is usually observed in the core of relaxed clusters. In the tail we find both low-entropy and high metal abundance (and also high pressure): this suggests that in the past this gas could have been at the center of a relaxed structure. Moreover,  this also indicates that the processes that are responsible for the formation of this substructure did not significantly change the entropy of the ICM and therefore they are close to adiabatic. \\
The results that come from the profiles are based on an underlying assumption: that the center from which we have extracted them (i.e. the peak of the surface brightness) is the center of the cluster. We have some indications that the point we chose could be the center of the potential well of A3558 (coincidence with the pressure peak ant the BCG), but we have already outlined that the center of the large scale symmetry does not coincide with this point (Sec. \ref{results}). However the ratio maps with the new center still show the presence of high pressure and low entropy gas in these regions. Moreover the presence of gas with temperature similar to that of the core does not depend on the choice of the symmetry.\\    

\subsection{The core}
The  metal abundance  profile of the cluster (Fig. \ref{met_tail}, red boxes) shows a peak at the center, where also the surface brightness is peaked and the temperature drops (except than in the SE direction). These three features are always found together in cool core clusters: in A3558, we find a similar situation in a cluster which is not  relaxed and which shows strong azimuthal variations in the thermodynamic quantities, especially in the SE sector.  However, all these features are not as apparent as in relaxed objects.\\
The deprojected central density is $n_e = (1.54 \pm 0.01) \cdot 10^{-2}\, \rm{cm}^{-3} $ and the cooling time in the inner 40 kpc is $t_{cool} \simeq 5 \cdot 10^9 \rm{yr} $, smaller than the Hubble time but larger than  the typical values found  in relaxed clusters \citep{peres98}

\section{Discussion}
Our analysis of the \xmmn and \chandra observations of A3558 has revealed many interesting characteristics.\\
 The cluster's dynamical history  is probably more complicated than expected, since it cannot be classified easily as a merging or relaxed cluster. It has some features similar to those of cool core clusters such as:
\begin{itemize}
\item An ``intermediate'' surface brightness peak at the center of the cluster (comparing Fig.~\ref{map_epic} with the flux images of a sample of nearby cluster we have found that A3558 surface brightness shows a clear peak at the center, while merging clusters do not, even if this peak is less apparent than in cool core clusters);
\item A peak at the center in the metal abundance profile, similar to those in cool core clusters \citep{grandi01}; in Figs.~\ref{cf_fig}d and \ref{met_tail},  we always find an increase of metalicity in the inner 60 kpc;
\item Presence of a BCG at the center and absence of other cD galaxies;
\item A small temperature drop  ($10-20\%$) at the center, except in the SE direction  (Fig.~\ref{map_ad}) where the temperature remains constant in the inner 200 kpc;
\item Cooling time smaller than the Hubble time in the inner 40 kpc. 
\end{itemize}
However, it also has other properties that are more common in   merging clusters:
\begin{itemize}
\item Significant deviations from spherical symmetry in the thermodynamic maps (as seen in  Figs. \ref{map_ad}, \ref{map_al} and also in the ratio maps in Fig. \ref{ratios}, the most deviating region is the SE sector where we measure low temperature, low entropy, high pressure and high density);
\item Low entropy tail (with high metal abundance) in the SE sector (Figs. \ref{map_ad} and \ref{map_al});
\item Pressure excess in the SE sector, detected with two independent methods (pressure ratio map, Fig.~\ref{ratios}, and deprojection, Fig.~ \ref{depro}).
\end{itemize}
Another puzzling feature of this cluster is the absence of extended radio emission \citep{venturi00}, that would be expected in massive merging clusters. \\
These ``strange'' characteristics of A3558 make the interpretation of its dynamical state a difficult task. From our data, we can exclude that the cluster is undergoing a major merger and that  it is relaxed.

\subsection{Origin of the anomalous SE regions}
In Sec.\ref{results} we discussed that the SE sector of A3558 has ``anomalous'' thermodynamic properties with respect to the other sectors: low temperature (similar to that in the core), low entropy, high pressure and high metal abundance. Therefore, these regions are  very important  to study the dynamical state of the cluster. In this section, we thus investigate different simple physical phenomena that could have lead to the formation of this anomalous structure.   

\subsubsection{Projection effects}
One of the simplest scenarios to explain the SE sector ``anomaly'' is to invoke projection effects on the main cluster gas. In this hypothesis the low temperature measured in the SE sector could be due to a cold substructure superposed on the line of sight of the cluster, while A3558 would have spherically symmetric distributions of the thermodynamic quantities and would look ``relaxed''.\\  
To test this hypothesis we fitted the spectra extracted in a circle in the ``tail'' (Fig. \ref{ratios}) with a 2--temperature model but the fit does not improve significantly and the temperature of the second component is very similar to that of the first. 
These results show that the proposed scenario cannot explain the observed data:
even if there were a second thermal component due to a projected substructure, the temperature of the ICM of the main cluster would remain lower than in the other sectors and we would still need another mechanism to explain this.\\ 
Since the anomaly is located in the infall region, where the filament connects A3558 to the other members of the core of the Shapley supercluster, a possible superposed substructure could be the filament itself. Therefore, we fitted the spectra with two thermal components, fixing the value of the second temperature in the range $1-1.5$ keV \citep{kull99}. Again the quality of the fit does not improve with this model and from the ratio of the normalizations of the components, we find that a  filament can contribute  less than $1\%$ to the emission of the region.

\subsubsection{An embedded subcluster}
\label{subcluster}
Since the pressure distribution is a tracer of the shape of the potential well of a cluster, the pressure excess in the SE sector may indicate the presence of the dark-matter halo of a small structure which is interacting with the main cluster. This scenario is consistent with the low temperature and entropy and the high metalicity measured in this region.\\
We have characterized the subcluster, using a model similar to the one described in Appendix B, assuming that region 2 in Fig. \ref{clump1} is a substructure completely filled by high density gas. Fitting the spectrum with a two-temperature model (one for the substructure and the other for the emission along the line of sight), we find that the substructure has temperature $kT=(3.82 \pm 0.07)$ keV and density $n=(2.54 \pm 0.03)\cdot 10^{-3} \, \rm{cm}^{-3}$ \footnote{The errors reported for temperature and entropy are only statistical.}, from which we can derive its gas mass $M_{gas}=3.89\cdot 10^{11}$ $ M_{\odot}$. Assuming  hydrostatic equilibrium, we have calculated the total mass contained in the pressure peak: it is in the range $(3.5-7)\cdot 10^{12}$ $M_{\odot}$, lower by a factor of 3-5 than the mass contained in the same radius at the center of the cluster. Since this model is based on strong assumptions, especially the choice of the surface brightness peak as the center of the cluster, we should treat the parameters that we have derived as rough estimates.\\
A mass $\sim 5\cdot10^{12} M_{\sun}$ is comparable with the mass of a large cD galaxy or of a small group, which usually have a virial temperature lower than the value that we derived for the subcluster. This discrepancy can be explained since we are not observing an isolated group, but we are describing the properties of a subcluster embedded in a massive cluster.
The subcluster has probably entered the main cluster from SE, from  the ``filament'' connecting A3558 to the SC group. This is the preferential direction of merging and the SE sector is therefore the infall region of matter in the potential well of the cluster.\\
There are two possible critical points of this model: the absence of correlation between the pressure excess and optical data and the lack of heating and visible hydrodynamic effects associated with the infall of the subcluster.\\
Comparing our pressure map and optical data \citep{bardelli98b}, we find no  clear indication of a particular concentration of galaxies corresponding to the pressure excess. However this is a weak objection since we do not expect this substructure to be associated with a large number of galaxies, because of its low mass. \\
The other objection is more serious: we would expect a merging subcluster to produce disturbances visible in the X-ray image and in the thermodynamic maps, while there is no evidence of heating in the infall region, which is furthermore the coldest sector of the cluster. Recently, \citet{ascas06} have shown a set of simulations where the merging between a massive cluster and even small subclusters produces global disturbances in the ICM visible in the X-ray image long after the first core passage. However, in another set of simulations, \citet{ascas06} show that if the  infalling subcluster has no gas there are no visible disturbances in the X-ray image, except the sloshing of the core (that we will discuss in Sec.\ref{or_cf}).
This could explain the lack of heating in the infall region of A3558: if the gas of an infalling subcluster has been completely stripped moving through the dense environment of the filament before reaching the cluster, we should treat the subcluster as ``dark-matter'' only and therefore we should not expect large hydrodynamic effects. However in this scenario, the subcluster would not contribute its low temperature and entropy ICM in the SE regions of the cluster. It can only perturb the gravitational potential and induce some hydrodynamic effects that could move the low-entropy gas from the core of the main cluster to the outer regions (in the case of A3558 up to 500 kpc from the surface brightness peak)  \citep{ascas06}.

\subsubsection{Stripping due to the motion of a cold cloud}
\label{stripping}
From the analysis of the cold front (Sec. \ref{coldfront}), we know that A3558 hosts a cloud of dense and cold gas which is currently moving NW. During its motion through the cluster it could have lost part of its high density and low temperature gas (``core-like'' ICM) because of the stripping from the surrounding environment. Indeed, the tail-shape of the low entropy region (Fig. \ref{map_ad}) suggests that some stripping has occurred. \\
This mechanism can be effective only if the ram pressure is large enough to overcome the internal pressure of the subcluster, i.e. in the condition 
\begin{equation}
\label{eq_rp}
\rho_{env}v^2 > P_{c}-P_{env},
\end{equation}
where $\rho_{env}$ is the density of the environment in which the cloud is moving with velocity $v$, $P_{c}$ is the internal pressure of the cloud and $P_{env}$ the pressure of the surrounding environment. We have calculated the velocity necessary to satisfy Eq. \ref{eq_rp} for a cold cloud that travels through the ICM of A3558:
\begin{equation}
\label{eq_speed}
v(r)=400 \rm{km}\,\rm{s}^{-1} \, \rm{keV}^{-1/2}\left(\frac{n_c T_c - n(r)T(r)}{n(r)}\right)^{1/2},
\end{equation}
where n(r) and T(r) are the properties of the ICM, as a function of the distance from the cluster center. In the hypothesis that the moving cloud can be described by the properties of the ICM inside the cold front, we derive that the required speed is consistent with the value obtained from the analysis of the cold front only at distances smaller than 150 kpc from the cluster center. The low entropy gas located at the end of the tail (500 kpc) cannot have been stripped and deposited ``in situ'' during the motion of the cloud through the cluster, since the stripping can be effective only in the inner regions where the density of the environment is large.\\

\subsubsection{A3558 former core}
\begin{figure*}
\vspace{-2.5 cm}
 \hspace {-2.5 cm}
   \includegraphics[width=22 cm, angle=90]{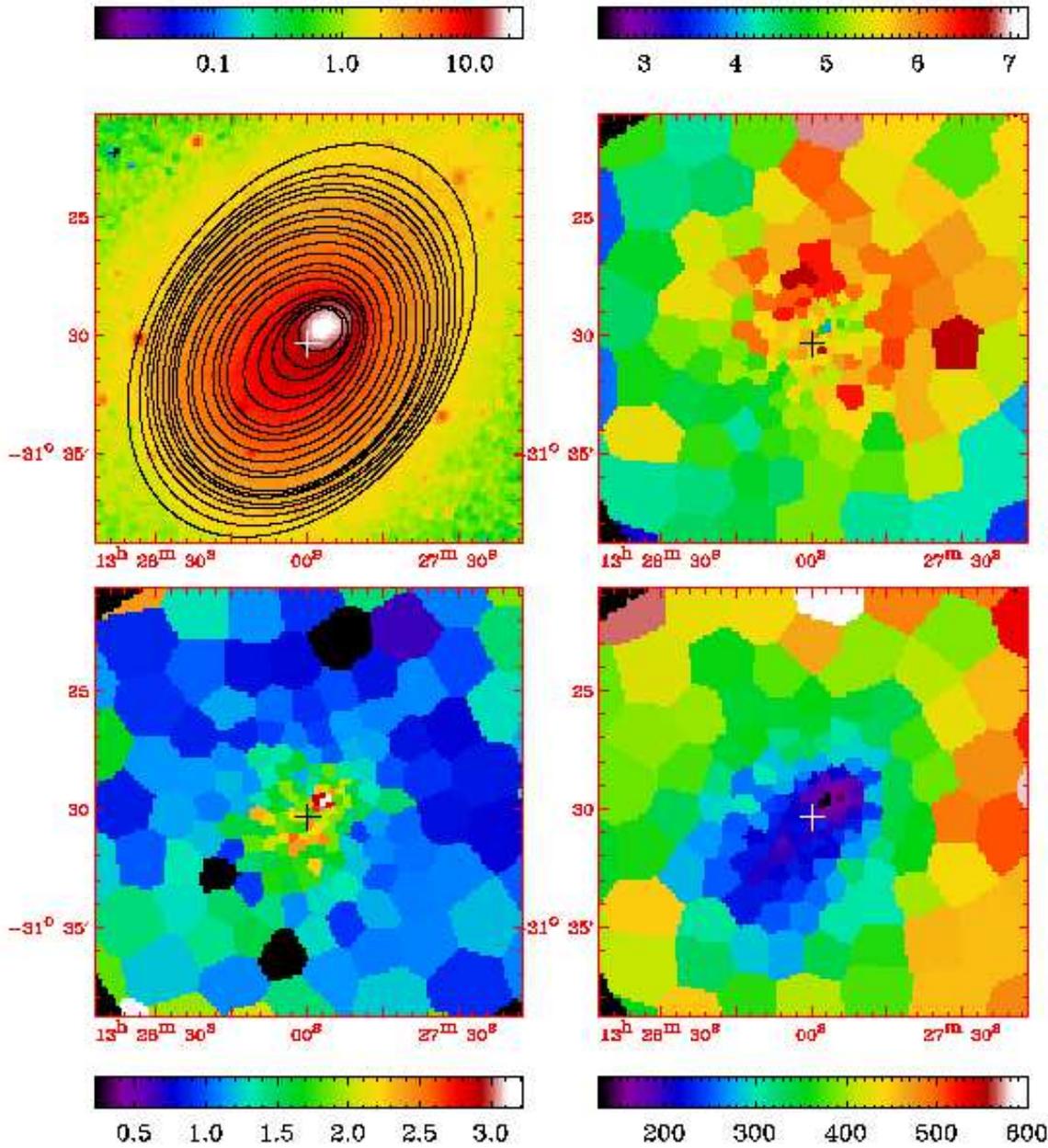}
   \caption{The crosses  show the position of the center of large scale symmetry. \emph{Upper left:} EPIC flux image with the ellipses used to approximate contours, \emph{Upper right:} temperature map ,   \emph{Lower left:} pressure ratio map and \emph{Lower right:}  pseudo-entropy map. The green cross shows the position of the centroid of X-ray large scale isophotes. }
              \label{centroid}%
    \end{figure*}

One interesting feature of A3558 is the fact that the center of the large scale isophotes does not coincide with the surface brightness and pressure main peak but is located $\sim 70$ arcsec SE of the BCG.  This effect has been observed in many clusters, but it is interesting that in A3558 the X-ray centroid is located in the pressure excess observed in the ratio map (Fig. \ref{centroid}). \\
This coincidence may have an interesting physical meaning: the large scale isophotal centroid could indicate the original position of the core before the processes that have lead to the formation of the cold front (Sec \ref{or_cf}).
We can assume that the center of the cluster was located at the actual position of the X-ray large-scale centroid, then that the core started to oscillate producing a cold front (\citealt{tittley05, ascas06}) and therefore distorting the isophotes in the inner regions, while the external part of the cluster is not perturbed. The low entropy, high pressure and high metal abundance gas located near the former center would be part of the ICM of the core.\\
The presence of low-temperature and -entropy gas SE of the surface brightness center (Fig. \ref{centroid}) can be explained only taking into account hydrodynamic processes. Figure 7 in \citet{ascas06} shows a similar situation even if at smaller scales: the ICM of the sloshing core is dragged outward by the motion of the surrounding gas.  
 
\subsubsection{Summary}
We have proposed several simple scenarios to explain the anomalous properties of the SE sector. Projection effects are ruled out by our analysis, the presence of an embedded subcluster requires very strong assumptions on the nature of the substructure and the stripping deposition in situ by an infalling subcluster would require higher velocities or densities. 
It is apparent from our analysis that all simple scenarios fail to explain the properties of the SE regions and that hydrodynamic processes, such as those described in \citet{ascas06}, should be taken into account.

\subsection{Origin of the cold front}
\label{or_cf}
In Sec. \ref{coldfront} we have shown that A3558 hosts a cold front, moving transonically with a  Mach number $M \simeq 0.8$. The metal abundance profile across the brightness discontinuity is consistent with a smooth decreasing profile and with a small ($30-40\%$) jump. \\
Several mechanism have been proposed in the literature for the origin of cold fronts: they could be the remnant of a merging subcluster, as in A3667 \citep{vikh01b}, or the core of the cluster ``sloshing'' in the potential well, as in A1795  \citep{mark01a, tittley05, ascas06}. \\
If the cold front that we have detected is the remnant of a subcluster that is falling on A3558 with a subsonic motion, we would expect to recognize also the parent cluster's core, and this is not clearly detected  either in the X-rays or in the optical. 
To support this hypothesis we have to make one of two strong assumptions: either that the moving subcluster is superposed on the line of sight to the core of the main cluster or that the main cluster core has been almost completely destroyed by the merging subcluster. 
Moreover, since we do not observe a metalicity discontinuity across the cold front, we also have to assume that the metal abundance distribution of the subcluster must be similar to that of the main  cluster. \\
The other possible origin of the cold front is the ``sloshing'' scenario, which can describe two phenomena: the oscillation of the gas component of the core \citep{mark01a} and the oscillation of the whole core, dark matter and gas in response to the off-axis merger between the cluster and a group, as suggested by \citet{tittley05}. Recently \citet{ascas06} have shown that these two cases are not separate, since a perturbation of the gravitational potential induces  motion of the gas, which quickly decouples from its dark matter component. However, we  compare our results also with the models of \citet{mark01a} and of \citet{tittley05}, because they predict simpler observational properties, which can be more easily compared with real data.\\
In the model of \citet{mark01a}, the gas has decoupled from its dark matter component and oscillates in the potential well of the cluster. Since in A3558 the pressure peak of the cluster (i.e. the bottom of the potential well) coincides with the peak of the surface brightness (i.e. the center of the gas cloud), as we can see in Fig. \ref{map_ad}, we conclude that, in this scenario, the moving gas cloud is now  in the center of its oscillation and therefore in the point of maximum velocity. This would also explain  the high Mach number ($M\simeq 0.8$), compared to other sloshing cold fronts \citep{mark01a}. \\
If the sloshing is due to the oscillation of the gravitational potential \citep{tittley05}, the coincidence between pressure and surface brightness is expected and we cannot determine the position of the oscillating body with respect to the center of its oscillation. In their paper, \citet{tittley05} highlight  three main observational signatures of this kind of cold front, but 
we could verify only the third (the gradients of compressed density must point through the center of the cluster) because we detected only one cold front. \\
The  environment of superclusters is ideal  to observe the sloshing class of cold fronts, since the merging rate is higher than in field clusters. We have identified two groups that may be responsible for the perturbations in the potential well: the ``subcluster'' (if we interpret the anomalous SE sector as a merging substructure as in Sec. \ref{subcluster}) and the group SC 1327-312. In either case, to be responsible for the oscillation of the core, these structures must have already passed the point of closest encounter with A3558. 
It is difficult to suppose that the low-mass subcluster has already reached the point of closest encounter with the core of A3558: we have to assume that it has entered with a non-zero angle with respect to the plane of the sky and it is now behind or in front of the cluster. In this scenario, the core would have already completed half of its oscillation. 
A past closest approach between A3558 and SC 1327-312 seems more plausible. The group  could have passed north of the cluster to reach its actual position. A similar scenario is proposed by \citet{bardelli98b}: they suggest that A3558 and another cluster (whose remnants are A3562, SC 1329-313 and SC 1327-312)   collided in the past. \\
It is difficult to compare our data more quantitatively with the model proposed by \citet{tittley05}, because their results are based only on two sets of simulations with fixed mass ratios (30:1 and 10:1) and none of our candidate groups has a similar mass ratio with respect to A3558 ($\sim 100:1$ for the subcluster and $\sim 4:1$ for SC 1327-312, \citealt{ettori97}). Moreover, as outlined by \citet{ascas06}, complex hydrodynamic processes should be taken into account.\\
Comparing our results with the possible cold front models does not allow us to rule out completely  any possible interpretation. However we prefer the ``sloshing'' model, since the interpretation of the cold front of A3558 as the remnant of a merging subcluster would require too much coincidence. 

\subsection{The global picture}
\label{wholesc}
In this section we attempt to develop a unified picture and to interpret the dynamical state of the cluster. We have excluded that the cluster is relaxed and we have shown that all the main features of the cluster are consequences of interaction. However, we still have to discriminate between two classes of merging scenarios: one where the observed features are a direct effect of the interaction and the other where we see the consequence of hydrodynamic processes that follow the perturbation of the gravitational potential.\\
A possible scenario is the infall of a subcluster, that has entered A3558 from SE and is currently moving towards NW. This merging will produce, as direct effects, the cold front (separation edge between the ICM of the subcluster and that of the main cluster) and the stripped tail. However, we calculated (Sec. \ref{stripping}) that a higher velocity would be required  for the stripping to form the tail. Moreover,  a subcluster falling in the potential well of A3558 would have acquired a kinetic energy much greater than the value that we derive from the velocity of the cold front. This ``missing energy'' should have been transformed into thermal energy, but we have no evidence of heating. This merger should have produced disturbances visible in the morphology of the cluster, while the presence of a large scale symmetry disfavors this scenario. We have already shown (Sec. \ref{or_cf}) that the interpretation of the cold front as a merging subcluster would require too much coincidence: the superposition between the subcluster and the core and similar metalicity.\\
The main features of A3558 can also be interpreted as an indirect effect of an interaction. The ``sloshing scenario'' can explain not only the cold front, but also the tail. As explained in \citet{ascas06} the cool gas would be displaced from the potential minimum and, because of the ram pressure of the surrounding gas, part of this cool gas would expand adiabatically and move outward.   This scenario is consistent with the observed large scale symmetry, whose center does not coincide with the surface brightness peak. Finally the absence of current strong interactions with other structures can explain the lack of extended radio emission and the presence of a sort of cool core.\\   
Our analysis favors a scenario where the gravitational potential of A3558 has recently been perturbed by a merger that has induced oscillations of the core without destroying  the structure of the cluster.
A qualitative comparison between our data and the simulations  shows that hydrodynamic effects, such as those described in \citet{ascas06}, may indeed play an important role in the physics of the ICM. 

\section{Conclusions}
Our analysis has shown that A3558 cannot be considered either a relaxed cluster, even if it has a sort of ``cool core'' in its inner regions, or a merging one. It hosts a cold front, moving NW, which is probably due to the oscillation of the gravitational potential. During these oscillations, hydrodynamic processes have produced a ``tail'' of low-entropy, high-pressure and metal rich ICM. The lack of current strong interactions with other structures can explain the absence of extended radio emission. \\
However, something has perturbed the dynamical state of A3558.  This ``perturbing factor'' is probably now outside A3558: a past merger with  SC 1327-312 or with a more massive cluster whose remnants are A3562, SC 1329-313 and SC 1327-312 (as suggested by \citealt{bardelli98b}), may have perturbed the gravitational potential.
 Indeed, in the processes related to the formation of a ``supercluster'' structure (\citealt{akimoto03}), there are frequent interactions that can have induced the ``sloshing'' of the core.\\

\section*{Appendix A: Fit of the SB profile across the cold front}
\begin{figure}
   {\includegraphics[width=8cm]{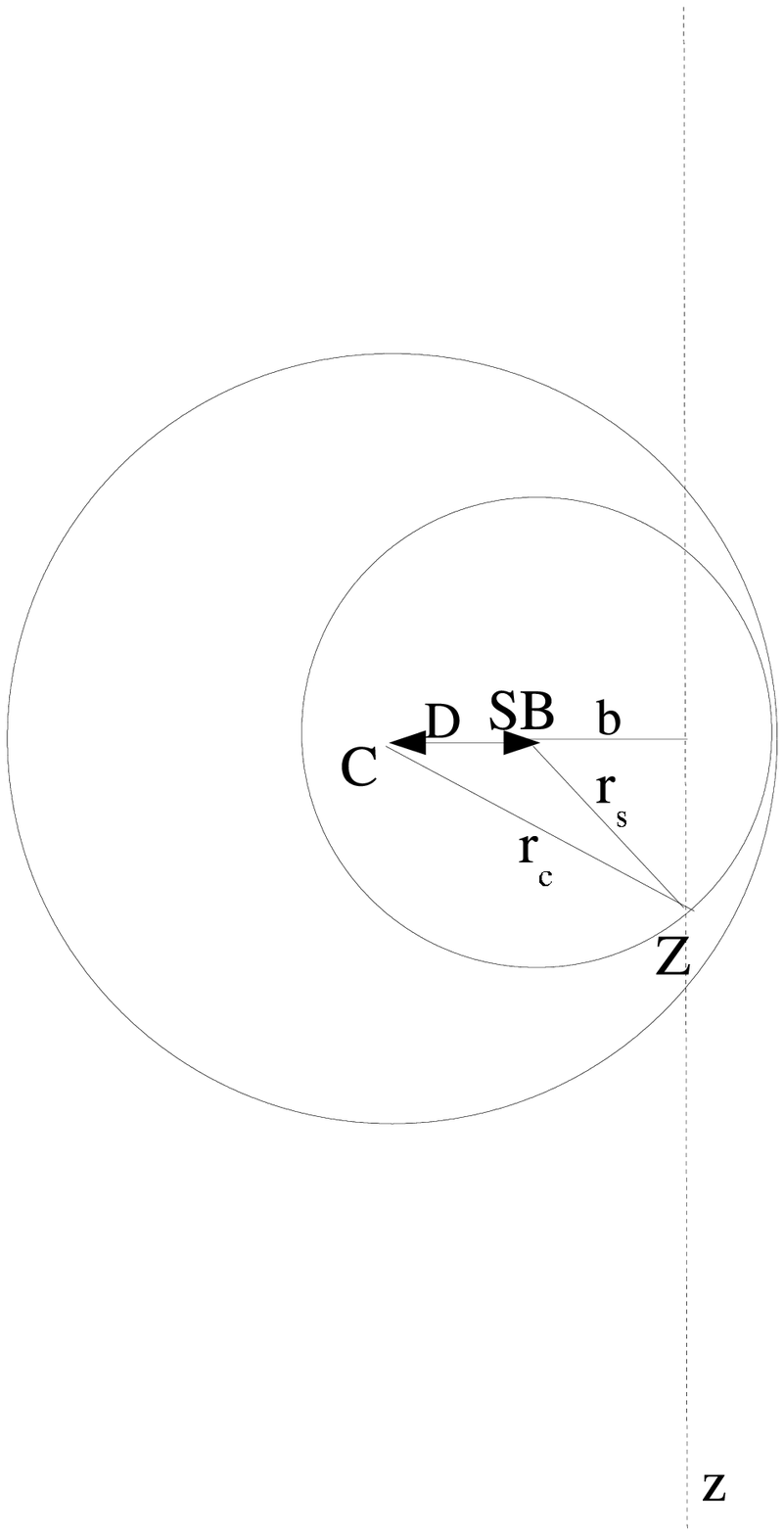}}
\caption{Geometrical model used to derive the density discontinuity across the cold front. C is the centroid of the surface brightness large scale contours, SB is the surface brightness peak, D is the distance between the centers, b is the projected distance, Z is the position along the z axis (perpendicular to the plane of the sky), $\rm{r}_s$ and $\rm{r}_c$ are the distances calculated from the surface brightness peak and the centroid, respectively.}
\label{modello}
\end{figure}
The density jump across the cold front can be derived from the surface brightness profile across the discontinuity.
 We assume the density profile to be composed of two $\beta$ models
\begin{equation}
n=\left\{\begin{array} {ll}
n_a\left(1+\left(\frac{r}{r_1}\right)^2\right)^{-\frac{3}{2}\alpha} & \rm{inside\, the\, cold\, front}\\ 
n_b\left(1+\left(\frac{r}{r_2}\right)^2\right)^{-\frac{3}{2}\beta} & \rm{outside} \end{array}\right.
\end{equation}
and we derive the parameters of this model ($n_a$, $r_1$, $\alpha$, $n_b$, $r_2$, $\beta$) and the projected distance of the edge $x_0$ (calculated from the brightness peak) from the fit of the surface brightness profile, which can be expressed as the integral of the emissivity along the line of sight:
\begin{equation}
\Sigma=KT^{1/2}\int_0^{+\infty}n^2(z)dz,
\end{equation}
where K is a constant. Since the dependence of the surface brightness on the temperature is weak, we do not consider its variation along the line of sight and we can leave it outside the integral. \\
The choice of the symmetry is very important: the profile inside the cold front should be centered on the center of curvature, which in our case coincides with the surface brightness peak, while the outer $\beta$ model should be centered on its own center of symmetry. From the analysis of the \xmmn observation (Sec. \ref{results}), we know that the large scale symmetry is elliptical and the center does not coincide with the surface brightness peak but is located $~70$ arcsec SE. This geometry is reproduced in Fig. \ref{modello}, where $z$ is the line of sight, $r_s$ and $r_c$ are the distances calculated from the surface brightness peak and the large scale symmetry center respectively and $D$ is the distance between the centers, that we assume equal to the projected distance $70$ arcsec.\\
To match the symmetry of the outer region, we have extracted the surface brightness profiles in four small elliptical sectors shown in Fig. \ref{cha_im}: in each of these sectors we can locally assume that the symmetry is spherical, and we can write the surface brightness as:
\begin{equation}
\Sigma=2B \int_0^{+\infty}\frac{dz}{\left(1+\left(\frac{r_c}{r_2}\right)^2\right)^{3\beta}},
\end{equation}
where $r_c$ is the distance calculated from the large scale center and $B=KT^{1/2}n_b^2$. Changing variables ($y^2=z^2/(r_2^2+b_c^2)$), this expression can be written as:
\begin{equation}
\Sigma=2Br_2 \left(1+\left(\frac{b_c}{r_2}\right)^2\right)^{-3\beta+\frac{1}{2}} \int_0^{+\infty}\frac{dy}{(1+y^2)^{3\beta}},
\label{rgtx0}
\end{equation}
where $b_c=b+D$ is the projected distance from the large scale symmetry center.\\ 
Since we are in the conditions $b_c >> r_2$ the expression in Eq. \ref{rgtx0} can be approximated by a power law $\Sigma=\bar{B}b_c^{-\gamma}$, where $\gamma=6\beta-1$. We fit the profiles and we derive $\gamma=1.36 \pm 0.09$ and the normalization.\\
The expression for the surface brightness inside the edge is more complicated since we must consider two components: one for the emissivity of the inner region integrated between  $0$ and $\sqrt{x_0^2-b^2}$ and the other for the emissivity of the outer region between  $\sqrt{x_0^2-b^2}$ and $\infty$:
\begin{eqnarray}
\Sigma & = & 2A\int_0^{\sqrt{x_0^2-b^2}}\frac{dz}{\left(1+\left(\frac{r_s}{r_1}\right)^2\right)^{3\alpha}}+{} \nonumber \\
& & {}+2B\int_{\sqrt{x_0^2-b^2}}^{\infty}\frac{dz}{\left(1+\left(\frac{r_c}{r_2}\right)^2\right)^{3\beta}},
\label{rltx0}
\end{eqnarray}
where $A=0.82KT^{1/2}n_a^2$ and  $r_s$ and $r_c$ are the distances calculated from the surface brightness peak and the large scale centroid respectively and $x_0$ is the curvature radius of the cold front. 
Under the same hypothesis that we have used in the previous case  ($b>>r_2$), the second part of Eq. \ref{rltx0} can be expressed as:
\begin{equation}
\Sigma_2 = \bar{B} \frac{\int_{\bar{y}}^{\infty}\frac{dy}{(1+y^2)^{3\beta}}}{\int_{0}^{\infty}\frac{dy}{(1+y^2)^{3\beta}}}(b+D)^{(-6\beta+1)}, 
\end{equation}
where $\bar{y}=\sqrt{x_0^2-b^2}/(b+D)$ and b is the projected distance from the surface brightness peak.\\
The first term in Eq. \ref{rltx0} cannot be approximated with a power law because we are not in the case $b>>r_1$:
\begin{displaymath}
\Sigma_1 = 2Ar_1\left(1+\left(\frac{b}{r_1}\right)^2\right)^{-3\alpha+\frac{1}{2}}\int_0^{\left(\frac{x_0^2-b^2}{r_1^2+b^2} \right)^{1/2}}\frac{dy}{(1+y^2)^{3\alpha}}.
\end{displaymath}
Fitting the surface brightness profile in the inner region, extracted in the spherical sector shown in Fig. \ref{cha_im}, with the function $\Sigma=\Sigma_1+\Sigma_2$, leaving the parameters of $\Sigma_2$ fixed to the best fit value for the outer region, we derive the parameters $n_a$, $r_1$ and $\alpha$.
Evaluating the $\beta$ model at $r=110$ we derive the density $n_{in}$ at the position of the discontinuity and the density jump $n_{in}/n_{out}=(1.43 \pm 0.04)$. The best fit model for the density is shown in Fig. \ref{cf_fig}b.

\section*{Appendix B: The clumping test}
To test if the observed excess could be due to clumping we have extracted the spectra in the region where the surface brightness excess is greatest (the white circle in Fig. \ref{ratios}), and fitted it with a two temperature model. To account for the projection effects we had to assume some geometrical and physical conditions.\\ 
\begin{figure}
   {\includegraphics[width=8cm]{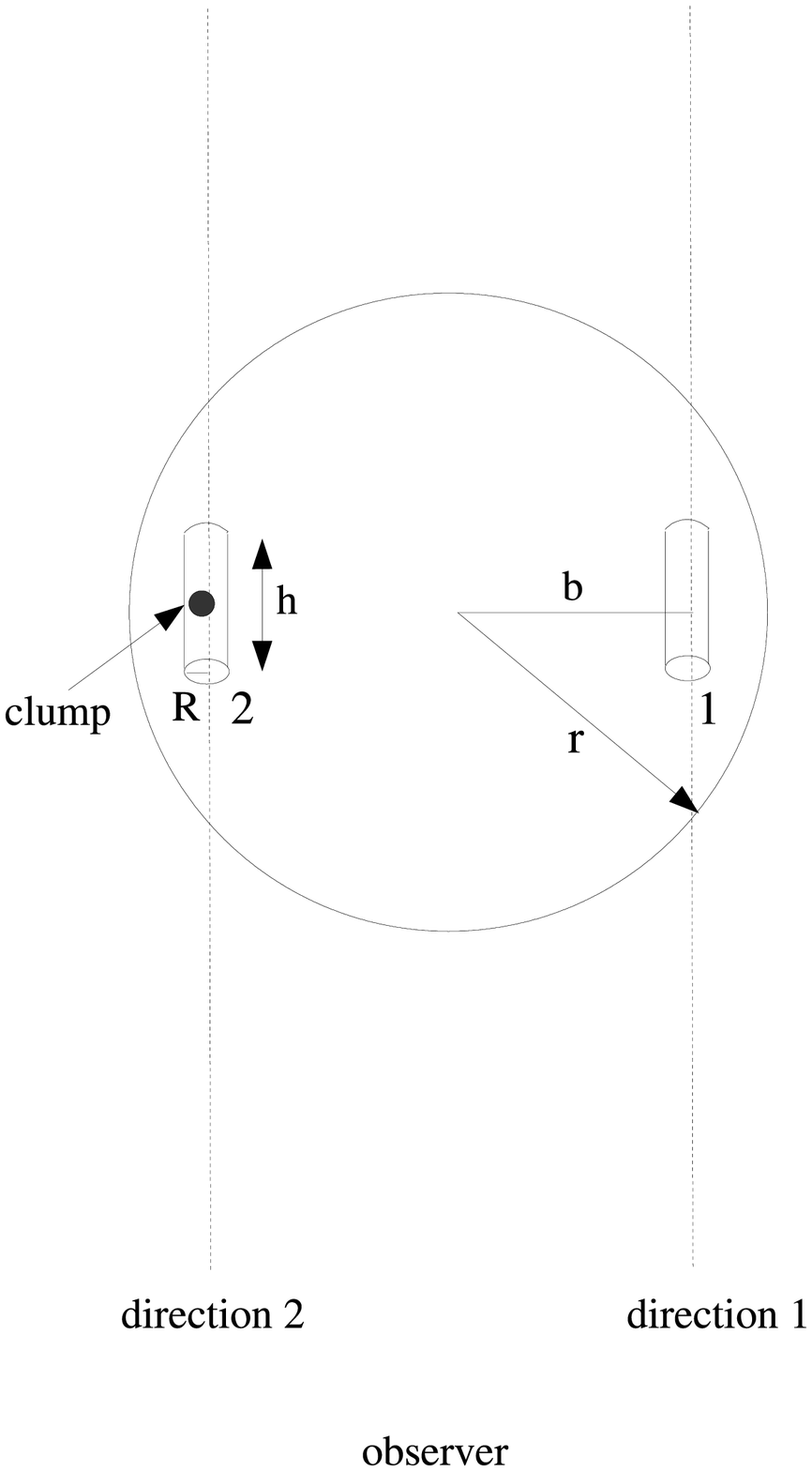}}
\caption{Geometrical model adopted for the clumping test.$\rm{R}$ is the radius of the circles from which we extract the spectra, $\rm{h}$ is the width of the region in which we assume there are clumps (along the line of sight), $\rm{b}$ and $\rm{r}$ are the projected and three-dimensional distances, respectively, calculated from the surface brightness peak.}
\label{clump1}
\end{figure}
We consider two identical cylinders (Fig. \ref{clump1}) of radius $R$ and height $h=2R$, with the bases perpendicular to the line of sight (the bases corresponds to the circles in Fig. \ref{ratios}: the white circle is the base of cylinder 2 while the black one is that of cylinder 1). We assume that the cluster is  spherically symmetric around the surface brightness peak , with the only dishomogenity located in cylinder 2 and filling its volume with a filling factor $\phi$.
The emission in direction 1 is due to the homogeneous gas with density $n_1$ of cylinder 1 and to the gas on the line of sight with density $n_e(r)$: the XSPEC normalization of the mekal model that we have used to fit the spectrum can be expressed as
\begin{equation}
N_1=2k\int_{h/2}^{\infty}n_e^2(r)Adr +kn_1^2V,
\end{equation}
 where $A$ is the area of the base, $V=Ah$ is the volume of the cylinder,  $k$ is the XSPEC constant
\begin{equation}
k=\frac{10^{-14}}{4\pi(D_a(1+z)^2)},
\end{equation}
and $D_a$ is the angular size distance to the source (cm). The value of the gas density at $b\simeq 200$ kpc,  $n_1=7.52\cdot10^{-4}\, \rm{cm}^{-3}$,  has been derived from the deprojection in spherical sectors.  \\
The emission of cylinder 2 has two thermal components: one due to the clump with density $n_c$ and the other due to the homogeneous gas with density $n_1$, filling the remaining volume. The normalization due to this second component ($\bar{N_2}$) can be expressed in terms of the normalization in direction 1, $N_1$: 
\begin{equation}
\bar{N_2}=2k\int_{h/2}^{\infty}n_e^2(r)Adr +kn_1^2(1-\phi)V= N_1 - kn_1^2\phi V.
\end{equation}
We have fitted the spectrum from direction 2, with a two-temperature model varying the temperature of the ``clump'' component and its filling factor. The best fit combination of these parameters, $T_c=3$ keV and $\phi=0.1$, leads to a density in the clump  which is about ten times higher than the density of the other component. Since the temperature ratio is much smaller ($T_u/T_c =1.97$), the clump component is not in pressure equilibrium with the surrounding medium. Even with greater filling factors the pressure equilibrium can never be achieved, unless we force the temperature to unrealistically low values which worsen significantly the quality of the fit. This analysis indicates that the pressure enhancement observed in the ratio maps is not an artifact  due to unresolved clumps and that it must have a physical origin.\\

\begin{acknowledgements}
We thank the anonymous referee for useful suggestions.
We thank M.Cappellari for providing the Voronoi 2D-binning method by Cappellari and Copin (2003) and for useful suggestions on the application of the method to X-ray data.
We thank S. Bardelli for providing optical data and S. Borgani for useful suggestions. M.R. is grateful for hospitality and support from  MPE in Garching. A.F. acknowledges support from BMBF/DLR under grant 50 OR 0207 and MPG, and is grateful for hospitality from INAF-IASF in Milan .
This paper is based on observations obtained with XMM-Newton, an ESA science
mission with instruments and contributions directly funded by ESA Member
States and the USA (NASA).
The
XMM-Newton project is supported by the Bundesministerium f\"{u}r Wirtschaft und
Technologie/Deutsches Zentrum f\"{u}r Luft- und Raumfahrt (BMWI/DLR, FKZ 50 OX
0001), the Max-Planck Society and the Heidenhain-Stiftung, and also by
PPARC, CEA, CNES, and ASI.
\end{acknowledgements}

\bibliographystyle{aa}
\bibliography{refs}

\end{document}